\documentclass[showkeys,showpacs,aps,amssymb,prd]{revtex4}

\usepackage{amssymb,latexsym,amsmath,amsthm,bm}
\usepackage{amsfonts}
\usepackage{amsbsy}
\usepackage{graphicx}
\usepackage{textcomp}
\usepackage{ifthen}
\newcounter{multi} \newcounter{multa}

\newcounter{faki} \newcounter{faka}

    \newtheorem{theorem}{Theorem}

    \newtheorem{corollary}[theorem]{Corollary}

\theoremstyle{definition} % For roman text in the body

\theoremstyle{remark} % For an italic header, more subtle than definition style

\newcommand{\be}{\begin{equation}}

\newcommand{\ee}{\end{equation}}

\newcommand{\beq}{\begin{eqnarray}}
\newcommand{\eeq}{\end{eqnarray}}

\newcommand{\beqa}{\begin{eqnarray}}
\newcommand{\eeqa}{\end{eqnarray}}

\newcommand{\beqn}{\begin{eqnarray*}}
\newcommand{\eeqn}{\end{eqnarray*}}

\newcommand{\beqan}{\begin{eqnarray*}}
\newcommand{\eeqan}{\end{eqnarray*}}

\newcommand{\comment}[1]{}

\newcommand{\R}{{\mathbb R}}

\newcommand{\one}{{\mathbf 1}}

\newcommand{\RR}{{\mathbb R}}

\newcommand{\ra}{\rightarrow}

\newcommand{\Jac}{\mbox{Jac} \,}

\def\frkS{{\mathfrak S}}

\begin{document}

\title{A Mathematical Theory of  Stochastic  Microlensing II.\\ Random Images, Shear, and the Kac-Rice Formula}

\author{A. O. Petters}
\email{petters@math.duke.edu} \affiliation{Departments of
Mathematics and Physics, Duke University, Science Drive, Durham,
NC 27708, United States of America}

\author{B. Rider}
\email{brider@euclid.colorado.edu} \affiliation{    Department of
 Mathematics,
    University of Colorado at Boulder,
    Campus Box 395
    Boulder, CO 80309,
    United States of America}

\author{A. M.
Teguia}\email{alberto@math.duke.edu} \affiliation{Department of
Mathematics, Duke University, Science Drive, Durham, NC 27708,
United States of America}

%\date{\today}

\begin{abstract}
Continuing our development of a mathematical theory of stochastic microlensing, we study the random shear and  expected number of random lensed images of different types. In particular, we characterize the first three leading terms in  the asymptotic expression of the joint probability density function (p.d.f.) of the random shear tensor due to point masses in the limit of an infinite  number of stars. Up to this order, the p.d.f. depends on the magnitude of the shear tensor, the optical depth, and the mean number of stars through a combination of radial position and the stars' masses. As a consequence, the p.d.f.s of the shear components are seen to converge, in the limit of an infinite number of stars,  to shifted Cauchy distributions, which shows  that the shear components have heavy tails in that limit. The asymptotic p.d.f. of the shear magnitude in the limit of an infinite number of stars is also presented. All the results on the random microlensing shear are given for a general point in the lens plane. Extending to general random distributions (not necessarily uniform) of the lenses, we employ the Kac-Rice formula and Morse theory to deduce general formulas for the expected total number of images and the expected number of saddle images. We further generalize these results by considering random sources defined on  a countable compact covering of the light source plane. This is done to introduce the notion of {\it global} expected number of positive parity images due to a general lensing map. Applying the result to microlensing, we calculate the asymptotic global expected number of minimum images in the limit of an infinite number of stars, where the stars are uniformly distributed. This global expectation is bounded, while the  global expected number of images and the global expected number of saddle images diverge as the order of the number of stars.

\end{abstract}



\maketitle

\section{Introduction}\label{section0}

We started an analytical study of stochastic microlensing in Paper
I (Petters, Rider, and  Teguia \cite{PRT}), where  we derived
stochastic properties of the random time delay function, lensing
map, and bending angle. We continue this work here by
investigating the expected number of critical points (lensed
images) of the time delay function, or equivalently, the expected
number of solutions of the random lens equation.

There is a fair amount of literature dealing with the non-random
number of images produced by a microlensing situation. One of the
first works on this subject showed that  a fixed source lensed by
a single  star can produce two distinct images (Einstein
\cite{einstein}). This result was  generalized to any finite
number of stars for both  single and multiple plane lensing, with
results on sharp  lower bounds and counting formulas  of the total
number  of images, the number of minimum images, number of saddles
images and number of maximum images (Petters
\cite{Petters1,Petters2,Petters3}; also see Chapters 11 and 12 of
the monograph by Petters, Levine, and Wambsganss \cite{Petters}).
Sharp upper bounds on the total number of images have also been
obtained (see S. H. Rhie \cite{Rhie} and Khavinson and Neumann
\cite{dmitri}).   The case of stars with random positions was
later examined using semi-analytical methods, yielding asymptotic
formulas in the limit of an infinite number of stars for the
expected number of minima (e.g., Wambsganss, Watt, and Schneider
\cite{wamb};  Granot, Schrecter, and Wambsganss \cite{GSW}).
However, these results are limited to first order approximations
and are not applicable to general random distributions of the star
positions or source positions.

In the current paper, we begin by deriving the first three 
terms in the asymptotic expansion of the microlensing shear tensor's   p.d.f.
in the limit of an infinite  number of stars. These results show that in such a limit, the
marginal p.d.f.s of the components of the random shear tensor
converge  in the sense of distributions  to shifted Cauchy
densities.
 We also show that the third term in this   asymptotic  p.d.f. of the shear
tensor depends on the square of the magnitude of the shear due
only to stars, the optical depth,  and  the
 mean number of stars. The asymptotic p.d.f. of the
magnitude of  shear is determined as well.

Next, we derive for a  lensing map (not necessarily due to
microlensing), a Kac-Rice type formula for the expected number of
positive parity lensed images (i.e.,   {\it minimum } and {\it
maximum} images) lying in a compact region of the lens plane. This
formula can be applied in a large number of random lensing
scenarios, and is obtained using the Kac-Rice technique for
computing the expected number of zeros of random mappings (e.g.,
Adler and Taylor \cite{Adlerjon};
 Azais and Wschebor \cite{ds1}; Forrester and Honner \cite{Forrester};
Sodin and Tsirelson \cite{Sodin,Sodin1,Sodin2}; Shepp and
Vanderbei \cite{Vander}). We then combine this formula and results
from Morse theory  to obtain additional general formulas for the
expected total number of images and the expected number of saddle
images.

The paper then concludes by applying the aforementioned Kac-Rice
formula and the asymptotic joint p.d.f. of the microlensing shear
tensor  to microlensing due to point masses with continuous matter
and shear. Here, we focus our attention on  the  ``global
expectation''
  of the number of minimum images (which will be defined later).
We consider the minimum images since numerical simulations have
shown that they are significant contributors to high magnification
events in microlensing light curves (e.g., Wambsganss, Watt, and
Schneider \cite{wamb}).

The outline of the paper is as follows: The basic lensing
framework is given in Section~\ref{section01}. We obtain the
 asymptotic joint p.d.f for the
microlensing shear components in Section~\ref{section1}.  In
Section~\ref{section2}, we derive general formulas for the
expected total number of positive parity images, the expected
total number of  images, and the expected total number of saddle
images.  In the same section, we define the  global expectation
and obtain an integral formula for the  global expected number of
positive parity images. Section~\ref{sec-average-min} presents
applications of these general formulas, along with the results on
the random microlensing shear, to determine a  global expected
number of minimum images in stochastic microlensing.

 \section{Basics}\label{section01}

A general gravitational lens system is governed by its time delay
function
$$
T_y (x) = \frac{|x-y|^2}{2} - \psi (x)
$$
and lensing map
$$
\eta (x) = x - \nabla \psi (x),
$$
where $y$ is the light source's position and $\psi$ is the
gravitational lens potential (Ref. \cite{Petters}, Chapter 6). A
{\it lensed image} of a light source at $y$ is a critical point
$x$ of  $T_y$, that is, a solution $x$ in $\RR^2$ of the equation
$\nabla T_y (x) = {\bf 0}$. It immediately follows that lensed
images are solutions $x$ in $\RR^2$ of the equation
$$
\eta (x) = y,
$$
which is called the {\it lens equation}.  When the potential
$\psi$ or the source position $y$ becomes random, the lens
equation has a random number of solutions (lensed images).

A lensed image $x$ is {\it nondegenerate}
 (resp., {\it degenerate}) if $x$ is a nondegenerate
 (resp., degenerate) critical point of $T_y$.
Using the Morse lemma (see Milnor \cite{Milnor}), we can classify
  a  nondegenerate light ray as either a {\it local minimum } (if $x$ has index 0),
a {\it local saddle }
  (if $x$ has index 1), or a {\it local maximum } (if $x$ has index 2)
  (see Ref. \cite{Petters}, p. 178 for more information).  The set of positive
parity lensed images  consists of all the minimum and maximum
lensed images.

In stochastic microlensing, we consider a  gravitational field due
to  a collection of $g$ stars
 at positions $\xi_1, \dots, \xi_g$ that are independent and identically
 distributed (i.i.d.) over a region of the lens plane. The
resulting  gravitational lens potential $\psi_g$, time delay
function $T_{y,g}$ at $y$, and lensing map $\eta_g$ become random
and are
 given by:
\beq
\label{defP}
\psi_g(x)&=&\frac{\kappa_c}{2} | x |^2 - \frac{\gamma}{2}(x_1^2-x_2^2) + \sum_{j=1}^g m_j  \log | x - \xi_j |,\nonumber\\
\label{defT} T_{y,g}(x) &=&   \frac{1}{2} | x - y |^2 -
\frac{\kappa_c}{2} | x |^2 +\frac{\gamma}{2}(x_1^2-x_2^2) - \sum_{j=1}^g m_j  \log | x - \xi_j |,\nonumber\\
 \eta_g(x)&=& \left ( (1 - \kappa_c + \gamma) x_1, (1 - \kappa_c + \gamma) x_1\right)
-  \sum_{j=1}^g m_j \frac{x - \xi_j}{|x - \xi_j|^2},\nonumber
\eeq
where $x = (x_1, x_2)$ is in $L = \R^2 - \{\xi_1, \dots, \xi_g\}$
(lens plane). Unless stated to the contrary, the light source
positions $y$,  the continuous matter density $\kappa_c$, external
shear $\gamma$, masses $m_j$, and number of stars
 $g$ are assumed fixed.   The star positions $\xi_j=(U_j,V_j)$
are random vectors. We shall consider the case of random light
source positions
 in Section~\ref{section2} and  Section~\ref{sec-average-min}.

 In Section~\ref{section1} and Section~\ref{sec-average-min}, we shall work  under the following
assumptions about microlensing:
\begin{itemize}

\item  Cartesian coordinates $(u,v)$ are assumed in the plane
$\R^2$ with $\bf 0$ as the  origin.

\item All masses are assumed equal: $m_1 = m_2 = \cdots = m_g
\equiv m\neq 0.$

\item The star positions $ \xi_i = (U_i, V_i)$ are independent and
uniformly distributed in  the disk $B({\bf 0},R)$ of radius $ R$
centered at ${\bf 0}$, that is,
\[
     (U_i, V_i)  \sim  {\rm Unif} \left\{ (u,v)\in \R^2 :  |u|^2 + |v|^2 \le R^2
 \right\},  \  \  \  \   \ i = 1, \cdots, g.
\]

\item The quantities $R$ and $g$ are related by the following
physical formula for  the  optical depth $\kappa_*$ in point
masses (stars):
\beq \label{kappa}
 \kappa_*&=& \frac{m~ g}{R^2}.
\eeq
The quantity $\kappa_*$ is fixed. For $x \in \R^2$, with $0\leq
|x| < R$, we shall call the combination  $\kappa_* |x|^2/m $ the
{\it mean number of stars} within the disc of radius $|x|$ about
the origin.
\end{itemize}

\noindent  {\bf Remarks:} The assumptions stated above provide a
good approximation  for several astrophysical scenarios. Note that
another lensing model  used in the literature is the elliptical
lens model (see Ref. \cite{Petters}, pp. 101-108 for more on lens
models).

\section{Random Shear Due to Point Masses}
\label{section1}

The {\it shear} at $x$ is a gravitational tug of stars on  the
light ray through $x$. The two components of the  shear tensor at
$x=(x_1,x_2)\in L$ due  to point masses (stars) is denoted by
$\Gamma_g^*(x)=(\Gamma_{1,g}^*(x),\Gamma_{2,g}^*(x))$,
 with components as  follows (see Ref. \cite{Petters}, p. 96):
 \vspace{0.1in}
$$
     \Gamma_{1,g}^*(x) =   \sum_{j=1}^g  \frac{ m( (U_j - x_1)^2 - (
 V_j
 - x_2)^2 )} { [ (U_j - x_1)^2 + (V_j - x_2)^2]^2 }
~~~{\rm and}~~~
  \Gamma_{2,g}^*(x) = \sum_{j=1}^g  \frac{ 2m(U_j - x_1) ( V_j -
 x_2) } { [ (U_j - x_1)^2 + (V_j - x_2)^2]^2 }.
$$
\vspace{0.08in}

We study the asymptotic behavior of the random shear
 tensor $(\Gamma_{1,g}^*,\Gamma_{2,g}^*)$ due to the point masses by
 obtaining several leading terms of the asymptotic  joint  p.d.f. of its components
 in the limit $g\ra \infty$. To do so, we assume
 (for the remainder of the paper) that the random shear tensor has
  an   absolutely continuous distribution function. We then obtain   the
following:

%\vspace{0.3in}

\begin{theorem}
\label{lemma1}   Let $x=(x_1,x_2)\in \R^2$ with $|x|<R$ be fixed.
The joint p.d.f. $f_{\Gamma_{1,g}^*(x),\Gamma_{2,g}^*(x)}$ of the
~ pair $(\Gamma_{1,g}^*(x), \Gamma_{2,g}^*(x))$ is given in the
 limit $g \ra \infty$ by:

\beq
\label{limdens1}
f_{\Gamma_{1,g}^*(x),\Gamma_{2,g}^*(x)}(z,w)&=&
\frac{\kappa_*}{2\pi(\kappa_{*}^2 + \Gamma^2)^{3/2}} \left[\ 1 + \
\frac{H_1(\Gamma)}{g}  \ + \ \frac{H_2(\Gamma;\,|x|)}{g^2}\right]
\ + \ O(g^{-3}),~~
\eeq
where $(z,w)$   denotes the possible values of $\,
(\Gamma_{1,g}^*(x),\Gamma_{2,g}^*(x))$, $\Gamma=\sqrt{z^2 \, + \,
w^2} \,$    denotes the possible values   of the shear tensor's
 magnitude, and
\begin{small}
\beq
\label{equation1}
H_1(z,w)
&=& \ \  H_1(\Gamma) \, \ \ \ = \ \kappa_*^2\,\frac{9\Gamma^2 - 6\kappa_*^2}{4(\kappa_*^2 + \Gamma^2)^{2}}.\\
&& \nonumber\\
 \label{equation2} H_2(z,w; |x|) &=& H_2(\Gamma;\, |x|)  =
\frac{\kappa_*|x|^2}{m}\frac{\kappa_*^2(6\kappa_*^2\, - \,
9\Gamma^2 )}{2(\kappa_*^2 +\Gamma^2)^{2}} - \frac{ \kappa_*^2(8
\kappa_*^4 - 24\kappa_*^2 \Gamma^2 + 3 \Gamma^4)}{4(\kappa_*^2 +
\Gamma^2)^3}
+ \frac{15 \kappa_*^4 \left(8 \kappa_*^4 - 40 \kappa_*^2\Gamma^2 + 15 \Gamma^4\right)}{32 (\kappa_*^2 +  \Gamma^2)^4}. \nonumber\\
\eeq
\end{small}
\noindent The p.d.f. at this order is a function of the magnitude
of the shear due only to stars, optical depth $\kappa_*$, and the
 mean number of stars within the disc of radius $|x|$ about
the origin.

\noindent The marginals of
$f_{\Gamma_{1,g}^*(x),\Gamma_{2,g}^*(x)}$ are
 $$
 f_{\Gamma_{i,g}(x)}(a_i)
=\frac{\kappa_*}{\pi (\kappa_*^2 + a_i^2)}\left[1 \, + \,
\frac{K_1(a_i^2)}{g}\, + \, \frac{K_2(a_i^2;\,|x|)}{g^2}\right] \
+ \ O(g^{-3}),\ \ i=1,2,
$$
where $a_1=z,\, a_2=w$ and
\beqn
K_1(h)&=&\kappa_*^2\,\frac{3 h - \kappa_*^2}{2(\kappa_*^2 + h)^2}.\\
K_2(h;\, |x|)&=& \frac{\kappa_* |x|^2}{ m}
\frac{\kappa_*^2(\kappa_*^2 - 3h)}{(\kappa_*^2 + h)^2} \, - \,
\frac{\kappa_*^2}{2} \frac{\kappa_*^4 - 6 \kappa_*^2 h +
h^2}{(\kappa_*^2 + h)^3} \, + \, \frac{3 \kappa_*^4}{4}
\frac{\kappa_*^4 - 10 \kappa_*^2 h + 5 h^2}{(\kappa_*^2 + h)^4}.
\eeqn

 \end{theorem}

\vspace{0.3in}

\begin{proof}[Proof:]We set  up coordinates in the same way as in
the proof of Proposition~1   in \cite{PRT}. For the reader's
convenience, we repeat the details. Let $x=(x_1,x_2)\in  \R^2$
with $ |x|<R$.
 Translate the rectangular coordinates $(u,v)$ so
that its origin ${\bf 0}$ moves to position $x$ and denote the
resulting new coordinates by $(u',v')$ and its origin by ${\bf
0}^{'}$.  The old origin ${\bf 0}$ in $(u',v')$ has coordinates
$(u',v')= -x$. Define $\omega_0$ to be the unique principal angle
with $\cos\omega_0=x_1/|x|$, $\sin\omega_0=x_2/|x|$, and let
$\omega \equiv\omega_0+ \pi$. Rotating the rectangular coordinates
$(u',v')$ counterclockwise by angle $\omega$ we obtain new
coordinates $(u^{\prime\prime},v^{\prime\prime})$ with origin
${\bf 0}^{\prime\prime}= {\bf 0}^{\prime}$.  The old origin $\bf
0$ now lies at position $(u^{\prime\prime},v^{\prime\prime}) =
(|x|,0) \equiv x^{\prime\prime}$ on the positive
$u^{\prime\prime}$-axis. Let $(\theta, r)$ denote polar
coordinates in the frame $(u^{\prime\prime},v^{\prime\prime})$.

Let $x=(x_1,x_2)\in  \R^2$ with $ |x|<R$ and set $u_x = u - x_1$,
$v_x = v-x_2$, and ${\mathfrak r}_x^2 = u_{x}^2 + v_{x}^2$. The
joint characteristic function of $(\Gamma_{1,g}^*(x),
\Gamma_{2,g}^*(x))$ is given by:
%
%\begin{small}
\beq
 \label{lastint}
 \left(
\varphi_{\Gamma_{1,g}^*(x),\Gamma_{2,g}^*(x)}(t_1, t_2)
\right)^{1/g} &=&\frac{1}{\pi R^2} \int_{B({\bf 0},R)}  \exp
\left[ i t_1  \left( \frac{m(u_x^2 - v_x^2)}{{\mathfrak r_x^4}}
\right)
  + i t_2 \left(   \frac{2mu_xv_x}{ {\mathfrak r_x^4}}  \right) \right]\, du dv \nonumber\\
& = &  \frac{1}{\pi R^2} \int_{ B(x,R-|x|)}
 \exp\left[ i\frac{m}{\mathfrak r_x^4} (t_1, t_2) \cdot ( u_x^2 - v_x^2,  2u_x v_x) \right] du dv\nonumber\\
&&\ \ \ + \ \frac{1}{\pi R^2}\int_{B({\bf 0},R)\backslash
B(x,R-|x|)}
 \exp\left[ i \frac{m}{\mathfrak r_x^4} (t_1, t_2)\cdot (  u_x^2 - v_x^2, 2u_x v_x ) \right] du dv. \nonumber\\
&=&  \frac{2}{R^2}   \int_0^{R-|x|} r J_0 \left( \frac{ m t}{
r^2}\right) \, dr
 \ + \ \frac{1}{\pi R^2}\int_{\Omega}
\exp\left[ \frac{i m t \cos\left(2(\theta + \omega')\right)}{r^2}   \right] d\theta dr, \nonumber\\
\eeq
%\end{small}
 with $t = \sqrt{t_1^2 + t_2^2}, \ 2\omega' = 2\omega + \phi, \ \phi=\arctan\frac{t_1}{t_2},\ $  $J_0$ is
 the zero Bessel function.

Now, with $\nu = m t$,  we get:
 \beq
\label{lastint2} \frac{2}{R^2}\int_0^{R-|x|} r J_0 \left(
\frac{\nu}{ r^2} \right)  dr &=&
\frac{\nu}{R^2}\left[\frac{(R-|x|)^2}{\nu}
{~}_1F_{2}(-\frac{1}{2};\frac{1}{2},1;\frac{-\nu^2}{4(R-|x|)^4}) \ - \ 1 \ \right]\nonumber\\
&=&-\frac{\nu}{R^2} \ + \ \frac{(R-|x|)^2}{R^2}\left[\ 1 \ + \
\frac{\nu^2}{4(R-|x|)^4} \ + \  O(R^{-8})\right], \eeq
%\end{small}
%
%
where   ${~}_1F_{2}$ is the hypergeometric function.

We use Taylor series expansion and term-by-term integration to
write the  second integral in equation~(\ref{lastint}) as:
%\begin{small}
 \beqn
\frac{1}{\pi R^2}\int_{\Omega} \exp\left[ \frac{i \nu
\cos\left(2(\theta + \omega')\right)}{r^2}   \right] du dv &=&
\frac{1}{\pi R^2} \left[ \ \pi (2|x|R \ - \ |x|^2) \ + \
\sum_{n=1}^{\infty} \frac{(i \nu)^n}{n!}\ I_n \ \right], \eeqn
%\end{small}
where, for integers $n\geq 1$,
\beqn I_n&=&
\int_{R-|x|}^{R +|x|} r \int_{-f(r)}^{f(r)} \frac{\cos^n\left(2(\theta + \omega')\right)}{r^{2n}} d\theta dr\\
&=& \int_{R-|x|}^{R +|x|}  \frac{1}{r^{2n-1}}  \int_{-f(r)}^{f(r)}
\left(a_n \ + \ \sum_{j=0}^{b_n} c_{n,j} \cos\left[2(n- 2j)(\theta + \omega')\right] \right) d\theta dr\\
&=& \int_{R-|x|}^{R +|x|}  \frac{1}{r^{2n-1}} \left(2 a_n f(r)  +
\sum_{j=1}^{b_n} \frac{c_{n,j} \cos\left[2(n-
2j)\omega'\right]}{n-2j} \sin\left[2(n- 2j)f(r)\right] \right)  dr
\ + \  \frac{c_{n,0} \cos\left[2n\omega'\right]}{n} \mathfrak{I}_{2n}
\eeqn
where
 \beqn
 \label{identity}
 \mathfrak{I}_n &=& \int_{R-|x|}^{R+|x|} \frac{\sin\left[n \, f(r)\right]}{r^{n-1}} \
 dr,
\eeqn
and the coefficients $a_n, b_n$ and  $c_{n,j}'$s are obtained by
finding the Fourier series of $\cos^n\theta$. We will prove in the
appendix that $\mathfrak{I}_n=0$ for every integer $n \geq 2$. We
then obtain:
\beqn
%I_0&=& \area{(\Omega)}\ = \ \pi (2|x|R \ - \ |x|^2),\\
I_1&=&  \ 0,\\
I_2 &=& \int_{R-|x|}^{R +|x|}  \frac{f(r) }{r^{3}}  dr
\ = \ \frac{\pi}{2}\frac{|x|(2R \ + \ |x|)}{(R^2 \ - \ |x|^2)^2},\\
I_3 &=&\frac{3\cos
2\omega'}{2}\int_{R-|x|}^{R+|x|}\frac{\cos(f(r))~\sin(f(r))}{r^5}\
dr
 \ = \ -\frac{3\cos 2\omega'}{2}\frac{\pi |x|^2R^2}{2(R^2 - |x|^2)^4} \ = \ O\left(R^{-6}\right),\\
|I_n| &\leq& \int_{R-|x|}^{R +|x|}  \frac{4\pi \ + \ 1}{r^{2n-1}}
dr \ \leq \ \frac{2 |x|\,(4\pi \ + \ 1)}{(R- |x|)^{2n-1}} \ = \
O\left(R^{-2n + 1}\right)\ \ {\rm for} \ n \geq \ 0 . \eeqn
Therefore,
 \beq
\label{lastint3} \frac{1}{\pi R^2}\int_{\Omega} \exp\left[ \frac{i
\nu \cos\left(2(\theta + \omega')\right)}{r^2}   \right] du dv &=&
\frac{2|x|}{R}-\frac{|x|^2}{R^2}   -  \frac{|x|\nu^2}{  2 R^5}   -
\frac{|x|^2\nu^2}{  4 R^6}
 -   \frac{|x|^3\nu^2}{   R^7}  +  O\left(R^{-8}\right).
\eeq

We now substitute equations  (\ref{lastint2}) and (\ref{lastint3})
into equation~(\ref{lastint}) to obtain
\beqn \left( \varphi_{\Gamma_{1,g}^*(x), \Gamma_{2,g}^*(x)}(t_1,
t_2) \right)^{1/g}
 &=& 1 - \frac{\nu}{R^2} + \frac{\nu^2}{4R^4} \ + \ \frac{|x|^2 \nu^2}{2 R^6} \, + \, O(R^{-8})\nonumber\\
&=&1 \, - \, \frac{\kappa_*t}{g} \, + \,
\frac{\kappa_*^{2}t^2}{4g^2}
 \, +  \,   \frac{|x|^2\kappa_*^3 t^2}{2 m g^{3}}  \,    +  \, O(g^{-4}).\nonumber
\eeqn

An asymptotic expansion of the characteristic function is then
 derived apply the logarithmic function of both sides of the
above equation:
\beqn \log \left(\varphi_{\Gamma_{1,g}^*(x),
\Gamma_{2,g}^*(x)}(t_1, t_2)\right) &=& g\log\left[ 1 \, - \,
\frac{\kappa_*t}{g} \, + \, \frac{\kappa_*^{2}t^2}{4g^2}
 \, +  \,   \frac{|x|^2\kappa_*^3 t^2}{2 m g^{3}}  \,    +  \, O(g^{-4})\right]\nonumber\\
&=&-\nu_1 \ - \ \frac{\nu_1^2}{4g} \ + \
\left(\frac{|x|^2\kappa_*\nu_1^2 }{2 m }\ - \
\frac{\nu_1^3}{12}\right)\frac{1}{g^2}
 \ + \   O(g^{-3}),
\eeqn
thus,
\beqn e^{\nu_1}\varphi_{\Gamma_{1,g}^*(x), \Gamma_{2,g}^*(x)}(t_1,
t_2) &=& \exp\left[- \, \frac{\nu_1^2}{4g}\, + \,
\left(\frac{|x|^2\kappa_*\nu_1^2 }{2 m }
\ - \ \frac{\nu_1^3}{12}\right)\frac{1}{g^2}  \ + \   O(g^{-3})\right] \nonumber\\
&=& \ 1 - \ \frac{\nu_1^2}{4g} \ + \
\left(\frac{|x|^2\kappa_*\nu_1^2 }{2 m } \ - \ \frac{\nu_1^3}{12}
\ + \ \frac{\nu_1^4}{32}\right)\frac{1}{g^2} \ + \   O(g^{-3}).
 \eeqn
Hence
 \beq
 \label{char} \varphi_{\Gamma_{1,g}^*(x),
\Gamma_{2,g}^*(x)}(t_1, t_2) &=& e^{\nu_1}\left[\ 1 - \
\frac{\nu_1^2}{4g} \ + \ \left(\frac{|x|^2\kappa_*\nu_1^2 }{2 m }
\ - \ \frac{\nu_1^3}{12} \ + \
\frac{\nu_1^4}{32}\right)\frac{1}{g^2}\right] \ + \   O(g^{-3}).
 \eeq

 By assumption, the joint p.d.f.   $f_{\Gamma_{1,g}^*(x),
\Gamma_{2,g}^*(x)}$ of $(\Gamma_{1,g}^*(x), \Gamma_{2,g}^*(x))$
exists, i.e., the Inverse Fourier Transform (I.F.T.) of the L.H.S.
of equation~(\ref{char}) exists. Consequently, so does the I.F.T.
of the R.H.S. of equation~(\ref{char}). The later  can be computed
as the sum of the I.F.T. of functions of the form  $p(g)\times
q(t_1,t_2)$ where $p(g)$ is a function of $g$ and
 $q(t_1,t_2)$ is independent
of $g$ and integrable.

Let
$$
\varphi_g (t_1,t_2)=e^{-\nu_1}\left[1 \, - \, \frac{\nu_1^2}{4g}\,
+ \, \left(\frac{|x|^2\kappa_*\nu_1^2 }{2 m } - \frac{\nu_1^3}{12}
+ \frac{\nu_1^4}{32} \right)\frac{1}{g^{2}} \right].
$$
We have
$$
f_{\Gamma_{1,g}^*(x), \Gamma_{2,g}^*(x)}(z,w)= ({\rm
IFT}(\varphi_g)) (z,w) +O(g^{-3}).
$$
Here  ${\rm IFT}(\varphi_g)$ is the I.F.T. of $\varphi_g$ and can
be obtained    by computing integrals of the form
\beq
\label{ift1}
\lefteqn{\frac{1}{(2\pi)^2}\int_{\R^2}e^{-\kappa_*\sqrt{t_1^2+t_2^2}}(t_1^2+t_2^2)^{n/2}
 e^{-i(t_1 z+t_2 w)} dt_1 dt_2}~~~~~~~~~~~~~~~~~~~~~~~~\nonumber\\
&=& \frac{1}{2\pi} (z^2+w^2)^{-n/2 - 1} \int_0^{\infty}r^{n+1}e^{-\kappa_* (z^2+w^2)^{-1/2}r} J_0(r) d r\nonumber\\
&=&  \frac{1}{2\pi}  \kappa_*^{-n-2}\Gamma(n+2)
{~}_2F_1\left(\frac{n + 2}{2} ,\frac{n + 3}{2} ; 1
;-\frac{z^2+w^2}{\kappa_*^2}\right).
\eeq

We then complete the proof of the first part of the theorem by
using equation~(\ref{ift1}) for $n=0, 2, 3, 4$.

  Now consider the following functions defined on $\R$:
\beqn
%\label{L1}
h_1(z)&\equiv& f_{\Gamma_{1,g}^*(x),\Gamma_{2,g}^*(x)}(z,w) \ \ \ {\rm for ~fixed }\ w \in \R. \\
%\label{L2}
h_2(w)&\equiv& f_{\Gamma_{1,g}^*(x),\Gamma_{2,g}^*(x)}(z,w) \ \ \ {\rm for ~fixed }\ z \in \R.  \\
%\label{L3}
k_1(z)&\equiv&  \frac{\kappa_{*}}{2\pi} \frac{1}{(\kappa_{*}^2 +
z^2 + w^2)^{3/2}}
\left[1 \ + \ \frac{H_1(z,w)}{g} \ + \ \frac{H_2(z, w; |x|)}{g^2}  \right]\ \ \ {\rm for ~fixed }\ w \in \R. \\
%\label{L4}
k_2(w)&\equiv&  \frac{\kappa_{*}}{2\pi} \frac{1}{(\kappa_{*}^2 +
z^2 + w^2)^{3/2}} \left[1 \ + \ \frac{H_1(z,w)}{g} \ + \
\frac{H_2(z, w; |x|)}{g^2} \right]\ \ \ {\rm for ~fixed }\ z  \in
\R.
\eeqn
We know that $h_1,\ h_2\in L^1(\R)$ since the random  shear's
 p.d.f. is integrable.
  We can then prove that  $k_1,\ k_2\in L^1(\R)$ by direct
computation. Hence, the second part of the theorem follows by
integration with respect to  $w$ and $z$.
\end{proof}

\vspace{0.3in}

\noindent {\bf Remark:}
 The leading function in the R.H.S. of equation~(\ref{limdens1}) becomes a
p.d.f. asymptotically in $g$.  In particular, set
$$
{\cal F}_{g,\,|x|}(z,w)= \frac{\kappa_*}{2\pi (\kappa_{*}^2 + z^2
+ w^2)^{3/2}}\left[1 \ + \ \frac{H_1(z, w)}{g} \ + \ \frac{H_2(z,
w;\,|x|)}{g^2}  \right].
$$
Then for  $g$ sufficiently large,
$$
{\cal F}_{g,\,|x|}(z,w)\geq 0.
$$
In addition,
$$
\int_{\R^2}{\cal F}_{g,\,|x|}(z,w) dz dw \ = \ 1
$$
for all $g \geq 1$ and $x \in \R^2 $.  An analogous result was
found in \cite{PRT} for the random lensing map (see  the
discussion after Corollary 10).

\vspace{0.3in}

\begin{corollary}
\label{corlemma1}
  For any $x\, \in \, \R^2$,
  the pair $(\Gamma_{1,g}^*(x), \Gamma_{2,g}^*(x))$
converges in  the sense of  distributions to  the pair
$(\Gamma^*_{1,\infty}, \Gamma^*_{2,\infty})$  with joint density
given by:
\beqn \label{limdens}
 f_{\Gamma^*_{1,\infty}, \Gamma^*_{2,\infty}}(z,w) = \frac{\kappa_*}{ 2\pi} \frac{1}{[\kappa_*^2+ z^2 + w^2]^{3/2}}
\eeqn
as $g\rightarrow \infty$.
Note that each marginal of $f_{\Gamma^*_{1,\infty},
\Gamma^*_{2,\infty}}$ is a {\rm(}stretched{\rm)} Cauchy
distribution:
$$
f_{\Gamma^*_{1,\infty}}(z)= \frac{\kappa_* }{\pi (\kappa_*^2 +
z^2)}
 ~~~{\it and}~~~
 f_{\Gamma^*_{2,\infty}}(w)= \frac{\kappa_* }{\pi (\kappa_*^2 +  w^2)}.
 $$
\end{corollary}

\vspace{0.3in}

Corollary~\ref{corlemma1} is consistent with the findings by
Nityananda and Ostriker \cite{NitOst84} and Schneider, Ehlers, and
Falco \cite{SchEhlFal92} (see p. 329). Note that the corollary
shows that the random shear components due to point masses have
heavy tails  as $g \rightarrow \infty$, an important point we have
not seen   emphasized   in the literature.

\vspace{0.3in}

\begin{corollary}
\label{corlemma2}
 Let   $x\in \R^2$ with $ |x|<R$ be fixed and consider the
magnitude of the shear at $x$ due only to
 stars, namely,
$\Gamma_g^*(x)=\sqrt{\left(\Gamma_{1,g}^*(x)\right)^2\ +\
 \left(\Gamma_{1,g}^*(x)\right)^2}.$
The  p.d.f. $f_{\Gamma_g^*(x)}$ of $\Gamma_g^*(x)$ is given in the
 limit $g \ra \infty$ by:
\beqn
%\label{shear-pdf}
f_{\Gamma_{g}^*(x)}(\Gamma)&=&\frac{\kappa_*\Gamma}{(\kappa_{*}^2
+ \Gamma^2)^{3/2}} \left[1\, + \, \frac{H_1(\Gamma)}{g}\ + \
\frac{H_2(\Gamma; \,|x|)}{g^2}\right] + O(g^{-3}).~~~~ \eeqn
Here $\Gamma$   denotes the possible values of $\Gamma_{g}^*(x).$
 \end{corollary}

\vspace{0.3in}

\begin{proof}[Proof:]
We use  the change of variable
\beqn ~&&\left\{
\begin{array}{lr} \Gamma_{1,g}^*(x)&= \ \Gamma_{g}^*(x)\cos\Theta(x)\\
\Gamma_{2,g}^*(x)&= \ \Gamma_{g}^*(x)\sin\Theta(x),
\end{array}
\right. \eeqn and obtain \beqn
f_{\Gamma_{g}^*(x),\Theta(x)}(\Gamma,\theta)
&=& \Gamma f_{\Gamma_{1,g}^*(x),\Gamma_{2,g}^*(x)}(\Gamma\cos\theta, \Gamma\sin\theta)\\
&=&  \frac{\Gamma}{2\pi}\frac{\kappa_{*}}{(\kappa_{*}^2 +
\Gamma^2)^{3/2}} \left[ \ 1 \ + \ \frac{H_1(\Gamma)}{g} +
\frac{H_2(\Gamma; |x|)}{g^2} \ \right] + O(g^{-3}). \eeqn
We can then integrate the equation above with respect to $\theta$
from $0$ to $2\pi$ and obtain the result stated in the Corollary.
\end{proof}

\newpage

\section{Expected Number of  Images: General Case}
\label{section2}

In the current section, we are not restricting ourselves to the
point-mass model as was the case in the previous section. We shall
determine results for the expected number of images for a {\it
general} lensing map.

\subsection{Expected Number of Positive Parity Images: General
Case and the Kac-Rice Formula}\label{subgen1}

  We work under the following broad, natural
assumptions and notation:
\begin{itemize}
\item  The gravitational field potential $\psi$ is an {\it
isolated} potential with a finite number of singularities, all of
which are infinite (see \cite{Petters}, p. 424 for more on
isolated potentials).

\item The potential $\psi$ and lensing map $\eta$ are smooth, except on a set of measure zero.

\item {\it Notation:}

\begin{itemize}
\item  For a set $A$, let \ \ $\one_{A}(x)=\left\{
\begin{array}{ll} 1&\ \ {\rm if }\ x\ \in \ A\\
0&\ \ {\rm if }\ x\ \notin \ A.
\end{array}
\right. $

\item  Set $ G(x)=\det ( \Jac\eta )(x)$,  \ $A=(0,\infty)$, and
${\cal G}_A=\{\nu\in\R^2:G(\nu)\in A\},$ where $\eta$ a general
lensing map.

\item Let $N_+=N_{min}+N_{max}$, which is the total number of
maximum and minimum images.

\item Let $N_+ (D,y) = $ total number of positive parity images
(minima and maxima) lying in a subset $D$ of the lens plane for
fixed source position $y$.

\end{itemize}
\end{itemize}

\noindent The determinant of the Jacobian of the general lensing map
$\eta$  is  given by (see \cite{Petters}, p. 182):
$$
 \det ( \Jac\eta )(x) =  (1-\kappa(x))^2 - \Gamma^2(x)
 =(1-\kappa(x))^2 -    \left[  \Gamma_{1}^2(x)+ \Gamma_{2}^2(x)
 \right].
$$
Here $\Gamma_{1}$ and $\Gamma_{2}$ are the shear components of the
general lensing map $\eta$ and are defined by
$$
\Gamma_1(x)=\frac{\psi_{11}(x)-\psi_{22}(x)}{2}~~~{\rm
and}~~~\Gamma_2(x)=\psi_{12}(x),
$$
 and $\kappa(x)$ is the convergence defined by
$$
\kappa(x)=\frac{\psi_{11}(x)+\psi_{22}(x)}{2},
$$
where $ \psi_{ij}=\partial^2\psi/\partial x_i\partial x_j. $ 

The main result is:

%\vspace{0.3in}

\begin{theorem}\label{Ricevariation}
{\rm \bf (Expected Number of Positive Parity Images)} Let
$D\subset \R^2$ be a closed disk.  Assume the following:
\begin{enumerate}
\item $P\left[\{x\in D:\eta(x)=y \ {\rm and} \  \det ( {\rm
Jac}\eta )(x)=0\}\right]=0.$ \item $\kappa$ is bounded on $L$ {\rm
({\it lens plane})}.
\end{enumerate}
Then for a test function $\rho:\R^2\ra \R$ with compact support,
we have
\beq \label{ricevar-gen1} \int_{\R^2}\rho(y)E[N_{+}(D,y)]dy&=&
\int_{\R^2}\rho(y) \int_D  E\left[\det (\Jac \eta)(x)\ \one_{{\cal
G}_A}(x) \Big| \eta (x) = y \right] \ f_{\eta (x)} (y) \ d x \ dy.
\eeq
and, for almost all $y$ in $\RR^2$,  the following Kac-Rice
formula holds for lensing:
\beq
\label{ricevar1}
 E[N_+ (D,y)] = \int_{D} E\left[ \left(
(1-\kappa(x))^2 -  \Gamma^2(x) \right)  \ \one_{{\cal G}_A}(x) \,
\Big|  \ \eta (x) = y \right] \,         f_{\eta (x)}  ( y )   \
dx.
\eeq

\end{theorem}

\vspace{0.3in}

Before we prove this theorem, it is important to note that  it can
be applied to a large range of lens models.   Moreover,  the
singularities form a set of measure zero and so contribute a value
of zero to these integrals (see Rudin \cite{Rudin}, p. 20).

The first assumption  of Theorem~\ref{Ricevariation} requires that
the random images are almost surely not critical points of the
random lensing map. The second assumption holds for optical depth
constant in $L$.

\vspace{0.1in}

\begin{proof}[Proof:] The proof  follows, with some modifications,
the proof of the  Kac-Rice formula  given by Azais
 and  Wschebor in \cite{ds1} (Theorem~2.1). A more general version of the Kac-Rice formula
 with a different proof is given by Adler and Taylor in \cite{Adlerjon}.
 The main tool used in the proof below is Federer's
co-area formula.

We start by fixing the source position at  $y$.
 By assumption~1
and results in \cite{Petters1,Petters2,dmitri,Rhie},  the random
function $N_{+}(D_{},y)$ is  uniformly bounded, except on sets in
$\R^2$ of measure zero. Let $\rho:\R^2\ra \R$ be a test function
with compact support. Since $\eta$ is  smooth almost everywhere in $\R^2$ and $\rho(y)$
is integrable, we can apply Federer's co-area formula (see Federer
\cite{Adlerjon}, p. 303) and obtain
$$
\int_{\R^2}\rho(y)N_{+}(D_{},y)dy=\int_{D_{}}|\det (\Jac
\eta)(x)|\rho(\eta(x)) \one_{{\cal G}_A}(x)dx.
$$
Taking expectations on both sides of the above equation yields
\beq
 \label{ricevarn1}
 \int_{\R^2}\rho(y)E[N_{+}(D_{},y)]dy= \int_{D_{}}E\left[\det (\Jac \eta)(x)\
\rho(\eta(x))\one_{{\cal G}_A}(x)\right]dx.
\eeq %%
 The L.H.S. (and hence the R.H.S.) of equation~(\ref{ricevarn1})
is bounded and the absolute value in the R.H.S. is dropped since
we are only considering positive parity images.  Using the basic
result
\beq
\label{basicresult}
 E[U] &=& \int_{\R^n} E[U|{\bf V} = {\bf v}] f_{\bf V} ({\bf v}) d{\bf v},
 \eeq
where $U$ is a real-valued random variable, ${\bf V}$ is a
$\R^n$-valued random variable, and ${\bf v}\in \R^n$, we obtain:
$$
E\left[\det (\Jac \eta)(x)\ \rho(\eta(x)) \one_{{\cal
G}_A}(x)\right] = \int_{\RR^2} E\left[\det (\Jac \eta)(x)\
\rho(\eta(x)) \one_{{\cal G}_A}(x) \Big| \eta (x) = y \right] \
f_{\eta (x)} (y) d y.
$$
 Note that by equation~(\ref{ricevarn1}), the expectation
 $E\left[\det (\Jac \eta)(x)\ \rho(\eta(x)) \one_{{\cal G}_A}(x)\right]$
 is bounded  on $D$, except on subsets of measure zero. The above equation then implies that
$$
\int_D \int_{\R^2} \left| E\left[\det (\Jac \eta)(x)\
\rho(\eta(x)) \one_{{\cal G}_A}(x) \Big| \eta (x) = y \right] \
f_{\eta (x)} (y) \right| dy dx
$$
 is finite.
Therefore:
 \beqn
 \int_{\R^2}\rho(y)E[N_{+}(D,y)]dy &=&\int_{D}
\int_{\RR^2}
E\left[\det (\Jac \eta)(x)\ \rho(\eta(x))\one_{{\cal G}_A}(x) \Big| \eta (x) = y \right] \ f_{\eta (x)} (y) d y d x.\\
&=&\int_{\RR^2} \int_D
E\left[\det (\Jac \eta)(x)\ \rho(\eta(x)) \one_{{\cal G}_A}(x) \Big| \eta (x) = y \right] \ f_{\eta (x)} (y)  dx d y \\
&=&\int_{\RR^2} \int_D
E\left[\det (\Jac \eta)(x)\ \rho(y) \one_{{\cal G}_A}(x) \Big| \eta (x) = y \right] \ f_{\eta (x)} (y)  dx d y \\
&=&\int_{\RR^2}  \rho(y) \int_D E\left[\det (\Jac \eta)(x)\
\one_{{\cal G}_A}(x) \Big| \eta (x) = y \right] \ f_{\eta (x)} (y)
\ d x \ dy, \eeqn
where the second equality follows from Fubini's theorem, the third
follows from the condition $\eta (x) = y$, and the last one
follows from the fact that $\rho(y)$ is deterministic and
independent of $x$. This proves the result in
equation~(\ref{ricevar-gen1}). Since the test function $\rho$ was
arbitrary,  equation~(\ref{ricevar1}) is valid  for almost all $y
\in \RR^2$.
\end{proof}

\vspace{0.3in}

{\bf Remark:}  For the critical case, if $\kappa(x)=1$ for almost
all (a.a.) $x\in L$, then $G(x)\notin A$ for a.a. $x\in L$, so
$\one_{{\cal G}_A}(x)=0$ for a.a. $x\in L$. Hence, $E(N_{+}(D, y))
=0$. This is consistent with results in Ref. \cite{Petters}.

\subsection{Expected Number of Images and Saddle Images: General
Case}\label{subgen3}

We can modify Theorem~\ref{Ricevariation} to obtain a formula for
the expected total number of images and the expected number of
saddle images lying in $D$. However, it is  simpler to use results
from Morse theory contained in the following theorem (see
\cite{Petters1,Petters2} and \cite{Petters}, Theorem 11.4, p.
424):

\vspace{0.3in}

\begin{theorem} {\rm \cite{Petters}} \label{petterstheo}
Let $T$ and $\eta$ be, respectively, a single-plane time delay
family and lensing map induced by an isolated gravitational lens
potential $\psi$. Suppose that $\psi$ has $g$ singularities, all
of which are infinite singularities. Then the total number of
lensed images of a light source at a non-caustic point $y$ is
finite and the following   holds:
\begin{enumerate}
 \item $N_{min}\geq 1,~~N_{sad}\geq g+N_{max}, ~~N_{sad}\geq N_{min}+g-1, N_+=N_{sad}-g+1.$
\item $N_{tot}=2N_{+}+g-1=2N_{sad}-g+1,~~N_{tot}\geq g+1$.
\end{enumerate}
\end{theorem}

\vspace{0.3in}

For a generalization of Theorem~\ref{petterstheo} to  the case
when the potential is not isolated, see Ref. \cite{Petters}, p.
422. We can apply Theorems~\ref{petterstheo} and
\ref{Ricevariation} to  obtain formulas for the expected total
number of images and  expected number of saddle images:

\vspace{0.3in}

\begin{corollary} \label{cor-Ntot-Nsad}
Let $D$ be a closed disk whose interior contains  the $g$
singularities of $\psi$. Then, under the assumptions of
Theorem~\ref{Ricevariation}, the following holds for almost all
$y$ in $\RR^2$:
 \beqn
 \label{version4}
E [ N_{tot}(D, y) ]  &=&
2\ \int_{D} E\left[\left((1-\kappa(x))^2 - \Gamma^2(x) \right)\ \one_{{\cal G}_A}(x) \, \Big| \ \eta (x) = y \right] \,f_{\eta (x)}( y )   \ dx \  + \ g \ - \ 1 \\
\label{version5} E [  N_{sad}(D, y)]&=& \int_{D} E\left[ \left(
(1-\kappa(x))^2 -  \Gamma^2(x) \right)  \ \one_{{\cal G}_A}(x) \,
\Big|  \ \eta (x) = y \right] \,
         f_{\eta (x)}  ( y )   \ dx \  + \ g \ - \ 1,
 \eeqn
 with self-evident notation.
 \end{corollary}

\vspace{0.3in}

\begin{proof}[Proof:]
Let $\xi_1,\cdots,\xi_g$ be the positions of the singularities.
Choose $\epsilon_0$ small enough so that no image lies inside
$B(\xi_j,\epsilon_0)$ for $j=1,\dots,g$ and
$B(\xi_j,\epsilon_0)\cap B(\xi_i,\epsilon_0)=\emptyset~$ for all
$i \neq j$. This  is possible since the total number of images is
bounded by
$$
g+1\leq N_{tot}\leq 5g-5
$$
(see \cite{Petters1,Petters2} for the lower bound and
\cite{dmitri,Rhie} for the upper bound).

Set $ D'=D\backslash \bigcup_{j=1}^gB(\xi_j,\epsilon_0). $ The
time delay function $T_y$ associated with $\psi$  generically
satisfies Morse boundary condition A relative to $D'$ (see
\cite{Petters}, p. 403). We can then apply
Theorem~\ref{petterstheo} to $D'$ and obtain
\beq \label{randomnum}
 N_{tot}'=2N_{+}'+g-1=2N_{sad}'-g+1,
 \eeq
where quantities with ``prime'' are related to $D'$.

Equation (\ref{randomnum}) is an equality of random variables, and
holds for almost all sources position $y\in \RR^2$. Therefore,
we can take the expectations and obtain:
\beqn \label{randomnum1}
E[N_{tot}']=2E[N_{+}']+g-1=2E[N_{sad}']-g+1. \eeqn
However, by construction,
\beqn N_{tot} (D,y) =N_{tot}',~~~~N_{+}(D,y) =N_{+}', ~~~~{\rm
and}~~~~ N_{sad}(D,y) =N_{sad}'. \eeqn
 The results then follow  from Theorem~\ref{Ricevariation}.
\end{proof}

\vspace{0.3in}

\noindent {\bf Remark:} In Corollary~\ref{cor-Ntot-Nsad}, the
expected  total number of images and number of saddles images both
depend  on the topology of the region $D\cap L$ since $g$ is its
first Betti number.

\subsection{Global Expectation: General Case}\label{subgen2}

The expectations we shall
study
are actually
generalizations of the expression $E[N_+(D,y)]$ in
equation~(\ref{ricevar1}) 
since the source position
is not known in the physical systems of interest
to us (e.g., microlensing).
   Intuitively speaking, the quantity
$E[N_+(D,y)]$ incorporates averaging with respect to the random
components of the lensing map $\eta$ for a fixed source positon
$y$. We shall extend this to averaging over random source
positions in a fixed  compact set with positive measure in the light source plane and
then generalize further to ``averaging'' over a family of such 
compact sets with positive measure.  When all these averages are carried out, we refer to
the result as a ``global expectation.'' We now make these ideas
precise.

The first extension of $E[N_+(D,y)]$ is to consider a light source
position in $\RR^2$ that is not fixed at $y$, but whose position
is a random vector $Y$ with density $f_Y$. We then define the
expected number of positive parity images in $D$ as follows:
\beq \label{average} E[N_+ (D, Y)] \equiv \int_{\RR^2}
E[N_+(D,y)]\ f_Y (y) dy. \eeq
When the p.d.f. $f_Y$ of $Y$ has compact support $\frkS_0$, we
write the p.d.f. as $f_{Y,\frkS_0}$ to emphasize the support and
denote the expectation (\ref{average}) by:
$$
E[N_+ (D, Y; \frkS_0)] \equiv \int_{\RR^2} E[N_+(D,y)]\
f_{Y,\frkS_0} (y) dy.
$$

We have:

%\vspace{0.3in}

\begin{corollary}
Let $Y$ be the random source position having a general p.d.f.
$f_{Y,\frkS_0}$ with compact support $\frkS_0$. Then, under the
assumptions of Theorem~\ref{Ricevariation}, we obtain:
 \beq
\label{version2}
 E[N_+ (D, Y;\frkS_0)] = \int_{D}  E \left[ \left(
(1-\kappa(x))^2 -  \Gamma^2(x) \right) \one_{{\cal G}_A}(x)  \
f_{Y,\frkS_0}( \eta (x))\, \right] dx.
\eeq
 \end{corollary}

\vspace{0.3in}

\begin{proof}[Proof:] The result follows from equation (\ref{ricevarn1}) with
$\rho = f_{Y,\frkS_0}$.
\end{proof}

\vspace{0.3in}

\noindent  {\bf Remark:}
 It is important to add that the general
formula in (\ref{version2}) does not require knowing the
distribution of the pair $(\Gamma_{1}, \Gamma_{2})$ {\em
conditioned} on $\eta = ( \eta_{1}, \eta_{2})$, as is the case for
equation~(\ref{ricevar1}). But equation (\ref{version2}) requires
knowledge of the joint distribution of the quadruple $(\Gamma_{1},
\Gamma_{2}, \eta_{1}, \eta_{2})$. However, as will be seen in
Section~\ref{sec-average-min}, knowledge  of the joint
distribution of the pair $(\Gamma_{1}, \Gamma_{2})$ will be enough
for a certain microlensing case of interest to us.

\vspace{0.3in}

\noindent {\bf Example.} Suppose that $Y$ is uniformly distributed
with p.d.f. $f_{Y,\frkS_0}$, where the compact support $\frkS_0$
is a subset  of the light source plane with nontrivial area:
$$
{\rm area}(\frkS_0) \equiv |\frkS_0| >0.
$$
Then:
$$
E[N_+ (D, Y; \frkS_0)] = \frac{1}{|{\mathfrak S}_0|}\int_{D}  E
\left[ \left( (1-\kappa(x))^2 - \Gamma^2(x) \right) \one_{{\cal
G}_A}(x) \ \one_{{\mathfrak S}_0}( \eta (x))\, \right] dx.
$$

\vspace{0.4in}

We are now ready to state the generalization of expectation from
the set $\frkS_0$ with positive measure  to a family of sets. Let $\{{\mathfrak S}\}$ be
a countable family of compact subsets with positive measure  covering the light source
plane $\R^2$. Suppose that each member ${\mathfrak S}$ has the
same area as a fixed member ${\mathfrak S}_0$. For each $\frkS$,
consider random source positions $Y$ that are uniformly
distributed with p.d.f. $f_{Y, {\mathfrak S}}$ compactly supported
in $\frkS$. Note that by construction we have:
$$
f_{Y, {\mathfrak S}}(y) \, = \,
\frac{\one_{{\frkS}}(y)}{|{\mathfrak S}|} \, = \,
\frac{\one_{{\mathfrak S}}(y)}{|{\frkS}_0|}.
$$
Now, the quantity $E[N_{+}(D, Y; {\mathfrak S})]$ captures an
average with respect to the randomness arising from the lensing
map $\eta$ as well as an average with respect to the random source
position $Y$ over $\frkS$. We now ``average'' $E[N_{+}(D, Y;
{\mathfrak S})]$ over the family $\{{\mathfrak S}\}$. Explicitly,
we define the {\it global expected number of positive parity
images} lying in $D$ for a uniform source by:
\beq
\label{globalmean00}
 \widehat{E}[N_{+}(D,Y;{\mathfrak S})]_{\{{\mathfrak S}\}}
 \equiv
\frac{1}{|{\mathfrak S}_0|} \ \int_{D} E\left[ \det \left(
\Jac\eta \right)(x) \, \one_{{\cal G}_A}(x)\right]  \ d x,
\eeq
where $|\frkS| = |\frkS_0|$. The  global expected number of
minimum images $\widehat{E}[N_{+}(D, Y; {\mathfrak
S})]_{\{{\mathfrak S}\}} $ is well-defined (i.e., bounded) since
$$
0\ \leq \det \left( \Jac\eta \right)(x) \, \one_{{\cal G}_A}(x) \
\leq 1
$$
 and $D$ is compact.
This definition captures the physical setup we need for the
microlensing application in Section~\ref{sec-average-min}. We
  shall  compute (\ref{globalmean00}) asymptotically to three orders
in the case of microlensing.

\section{Global Expected number of images: Microlensing case}
\label{sec-average-min}

We now consider the  case of microlensing,  which consists of
randomly distributed point masses with continuous matter and
shear, as described in Section~\ref{section01}.    We shall apply
the results given in Sections~\ref{section01} and~\ref{section2}
to study the {\it global } expected number of  minimum images in
microlensing.

Consider a closed disk $D$.  The random   lensing map $\eta_g$
satisfies
 the assumptions made at the start of the
previous section and in the statement of
Theorem~\ref{Ricevariation}. In fact,
$$
P\left[\{x\in D:\eta_g(x)=y~~{\rm and}~~\det ( \Jac\eta_g
)(x)=0\}\right]=0,
$$
 (i.e., lensed images in $D$ are almost surely  not images of
caustic points) and
$$
P[\partial D\cap \{x\in D:\eta(x)=y\}=\emptyset]=1
$$
(i.e., there is almost surely  no image on the boundary of $D$)
since there are finitely many lensed images for almost all source
positions $y$ (see \cite{Petters}, p. 431 for more details). Note
that for microlensing, we have
$$
\Gamma_1 (x) =\gamma + \Gamma_{1,g}^*
(x),~~~~~~~~\Gamma_2(x)=\Gamma_{2,g}^* (x), ~~~~{\rm and}~~~~
\kappa(x)=\kappa_c \ \ {\rm for} \ x \in L,
$$
where $L = \RR^2 - \{\xi_1,\dots,\xi_g  \}$ (lens plane), and that
$D$ may contain singularities $\xi_i$.

Assume that $\kappa_c<1$ (subcritical) in the current section.
This implies that $N_{max}=0$ (see \cite{Petters}, p. 433).
Consequently, for the microlensing map $\eta_g$ with $\kappa_c
<1$, we have the following  for any non-caustic source position
$y$:
$$
N_+ (D,y) = N_{min} (D,y) \equiv N_{g,min} (D,y).
$$
Furthermore, the set ${\cal G}_A=\{\nu\in \R^2: \det (\Jac
\eta_g)(\nu) \in (0,\infty) \}$,  when $\kappa_c<1$, becomes:
%
%\begin{small}
\beqn
{\cal G}_A &=& \left\{\nu\in \R^2: \left(\gamma +
\Gamma_{1,g}^* (\nu) \right)^2
+  \left(\Gamma_{2,g}^* (\nu) \right)^2 <  (1- \kappa_c)^2 \right\} \\
&=& \left\{\nu\in \R^2: \left(\Gamma_{1,g}^* (\nu), \Gamma_{2,g}^*
(\nu) \right)  \in B((-\gamma, 0), 1 - \kappa_c)\right\}.
\eeqn
%\end{small}

We are interested in determining the global expected number of
minimum images in $D$, namely, we shall compute
(\ref{globalmean00}) for microlensing. By (\ref{globalmean00}),
the global expected number of minimum images would then be
obtained   by averaging over the possible uniform star positions
(i.e., the randomness arising from the lensing map), followed by
averaging the source position over a compact set with positive measure $\frkS$,  and
then ``averaging'' over the possible sets in $\{\frkS\}$, where
each has the same area.

We obtain:

\vspace{0.3in}

\begin{theorem} \label{theo}
Let $D\subset\R^2$ be compact and suppose that $\kappa_c<1$. Then
the  global mean number of   minimum images lying inside $D$  is
given by:

\begin{small}
\beq \label{eqtheo}
 \widehat{E}[N_{g,min}(D, Y; {\mathfrak
S})]_{\{{\mathfrak S}\}}
 &=&\frac{\kappa_* \ \mu_{D,{\mathfrak
S}_0}}{2\pi} \int_{{\cal B}} \frac{(1-\kappa_c)^2 - (\gamma+z)^2 -
w^2  }{(\kappa_{*}^2 + z^2 + w^2)^{3/2}} \left[ 1   +
\frac{H_1(z,w) }{g}   +  \frac{H_3(z,w) }{g^2}   \right] dzdw  +
O(g^{-3}),\nonumber \\
\eeq
\end{small}
in the large  $g$ limit, where
\begin{small}
$$
{\cal B}=B\left((-\gamma,0), 1-\kappa_c\right), \qquad
\mu_{D,{\mathfrak S}_0}=\frac{|D|}{|{\mathfrak S}_0|}, \qquad
H_3(z, w)= H_2(z, w; a_0), \qquad a_0\equiv \frac{1}{|D|}\int_D
|x|^2 dx,
$$
\end{small}
   and $H_1$ and $H_2$ are  defined in equations~(\ref{equation1}) and
   (\ref{equation2}).
Recall that $|{\mathfrak S}|\, = \, |{\mathfrak S}_0|$ by
definition of the global expectation.
\end{theorem}

\vspace{0.3in}

Note that $\widehat{E}[N_{g,min}(D, Y; {\mathfrak
S})]_{\{{\mathfrak S}\}}$ depends on the common mass $m$ since
$H_2$ is a function of $m$. Also  the integral on the R.H.S. of
equation(\ref{eqtheo}) remains bounded independently of $D$
although $a_0$ can be arbitrarily large.

%\vspace{0.3in}

\begin{proof}[Proof:]

By definition,
\beqn \widehat{E}[N_{g,min}(D, Y; {\mathfrak S})]_{\{{\mathfrak
S}\}} \ = \ \frac{1}{|{\mathfrak S}_0|} \ \int_{D} E\left[ \det
\left( \Jac\eta_g \right)(x) \,  \one_{{\cal G}_A}(x)\right]  \ d
x. \eeqn
Using Theorem~\ref{lemma1},  the integrand in the R.H.S. of the above
  equation,   which is bounded, can be written
in the microlensing case as:
\begin{small}
\beqn E\left[ \det \left( \Jac\eta_g \right)(x) \,  \one_{{\cal
G}_A}\right] &=& E \left[ \left((1 - \kappa_c)^2 - (\gamma +
\Gamma_{1,g}^*(x))^2 - (\Gamma_{1,g}^*)^2(x) \right)
\one_{{\cal G}_A}(x) \right]\\
& = &\int_{{\cal B}} \left((1-\kappa_c)^2 - (\gamma+z)^2 - w^2  \right) f_{\Gamma_{1,g}^*(x), \Gamma_{2,g}^*(x)}(z,w)dzdw\\
& =&\frac{\kappa_{*}}{2\pi}\int_{{\cal B}} \frac{(1-\kappa_c)^2
-(\gamma+z)^2 - w^2  }{(\kappa_{*}^2 + z^2  + w^2)^{3/2}} \left[1
+ \frac{H_1(\Gamma)}{g}   +  \frac{H_2(\Gamma; |x|)}{g^2}
\right]dzdw \ + \ O(g^{-3}), \eeqn
\end{small}
for large $g$.  Integrating over $D$ then yields the desired
result.
\end{proof}

\vspace{0.3in}

A physical assumption often used by astronomers is the following:
On the macro-lensing scale (e.g., galaxy scale), the macro-lens
potential is averaged over the stars' positions and the lensing map
is modeled by the transformation $\eta_{macro}:D\ra \R^2$
 given by
$$
\eta_{macro}(x_1,x_2)=((1-\kappa_{tot}+\gamma)x_1,(1 -
\kappa_{tot} - \gamma)x_2),
$$
where $\kappa_{tot}=\kappa_* + \kappa_c$. In this case, choosing
${\mathfrak S}_0 = \eta_{macro}(D)$ gives
$$
\mu_{D,{\mathfrak S}_0}=\frac{|D|}{|{\mathfrak S}_0|} =
\frac{1}{|(1 - \kappa_{tot})^2 + \gamma^2|} \equiv \mu_{macro},
$$
which we call the {\it macro-average total magnification } over
$D$.
 For the case when ${\mathfrak S}_0 = \eta_{macro}(D)$, we define:
 \beqn
%\label{eqexpD}
\widehat{E}[N_{g,min}(D)]&\equiv& \widehat{E}[N_{g,min}(D,Y;
{\mathfrak S})]_{\{{\mathfrak S}\}}.
 \eeqn

\noindent Recall that the stars   positions are uniformly distributed in the
disc  $B({\bf 0},R)$ of radius $R = \sqrt{m g/\kappa_*}$.

\vspace{0.3in}

\begin{corollary}\label{cor5}
 Let $D_r$ be a closed disc  of radius $r$ .
Assume that ${\mathfrak S}_0 = \eta_{macro} (D_r)$. The {\it
global mean number of minimum images} is:
\beqa
 \label{eqcor5}
\widehat{E}[N_{min}]&\equiv& \lim_{r \ra \infty
}\lim_{g\ra\infty}\widehat{E}[N_{g,min}(D_r)] =
\frac{\kappa_{*}}{2\pi\vline(1-\kappa_{tot})^2-\gamma^2\vline}
  \int_{{\cal B}} \frac{(1-\kappa_c)^2 -(\gamma+z)^2 -w^2}{ (\kappa_{*}^2+ z^2 + w^2)^{3/2} }  \, dz
  dw,\nonumber \\
& & 
\eeqa
and the global mean number of images and the global mean
number of saddle images grow as order $O(g)$ for large $g$,
with self-evident definitions.
\end{corollary}

\vspace{0.3in}

\begin{proof}[Proof:] We have $\mu_{D,{\mathfrak S}_0} = \mu_{macro}$.
 The
functions $H_1$ and $H_2$  defined by equation (\ref{equation1})
are bounded over $\R^2$ and $\R^2\times [0,r]$ respectively for a
fixed $r$. Thus, the function $H_3$ is bounded, independently of
$g$. Therefore, the integrand of the integral term in
equation~(\ref{eqtheo}) is bounded   when we substitute $D_r$ for
$D$. We can then take the limit as $g\ra \infty$ in equation
(\ref{eqtheo}) with $D$ replaced by $D_r$ and apply  the Dominated
Convergence Theorem to obtain
\[
 \lim_{g\ra\infty} \widehat{E}[N_{g,min}(D_r)]=
\frac{\kappa_{*}}{2\pi\vline(1 - \kappa_{tot})^2 - \gamma^2\vline}
  \int_{{\cal B}} \frac{(1 - \kappa_c)^2 - (\gamma + z)^2 - w^2}{ (\kappa_{*}^2 + z^2 + w^2)^{3/2} }  \, dz   dw,
\]
which is independent of $D_r$. Taking the limit as $r \ra \infty$
yields the first result of the corollary. The second  one  then
follows from Corollary~\ref{cor-Ntot-Nsad}.
\end{proof}

\vspace{0.3in}

\noindent {\bf Remark:}
\begin{enumerate}
\item For physically relevant values of $\kappa_*,\kappa_c$ and
$\gamma$, the R.H.S. of (\ref{eqcor5}) is finite.  However, the
global mean  is divergent for $(1-\kappa_{tot})^2=\gamma^2$.

\item This corollary agrees with the findings  by Wambsganss,
Witt, and Schneider \cite{wamb} and Granot, Schrecter, and
Wambsganss \cite{GSW}. The integral in equation~(\ref{eqcor5}) was
resolved  in closed form using elliptic integrals (see  Ref.
\cite{GSW}).
  \end{enumerate}

In the table below, we illustrate $\widehat {E} [N_{min}]$ for
$\kappa_{tot} = \kappa_c + \kappa_* = 0.450$ and varying $\gamma$
and $\kappa_*$. For all the values of $\kappa_*$ and $\gamma$
shown,
 we have the ``macro-minimum'' situation
of  $1- \kappa_{tot} + \gamma >0 $ and  $1- \kappa_{tot} - \gamma
>0$. Note that the number of images increases significantly when
$\gamma$  gets close to $1-\kappa_{tot}$.  In the column for
 $\kappa_* = 0.45 = \kappa_{tot}$, note how
${\widehat {E}} [N_{min}]$ increases to $6.108$ as  $\gamma$
increases.

%\vspace{0.3in}

\begin{center}
\begin{tabular}{|c|c|c|c|c|c|}\hline
\multicolumn{6}{|c|}{} \\
 \multicolumn{6}{|c|}{${\widehat {E}} [N_{min}]$:\ Global Mean Number of Minima } \\
 \multicolumn{6}{|c|}{} \\
 \cline{1-6}
$ \gamma\diagdown\kappa_*$&~~~0.1~~~&~~~0.2~~~&~~~0.3~~~&~~~0.4~~~&~~~0.45~~~ \\
\hline
   0.1& ~~~1.037~~~&~~~1.114~~~&~~~1.219~~~&~~~1.345~~~&~~~1.415~~~ \\
   0.2& ~~~1.068~~~&~~~1.171~~~&~~~1.230~~~&~~~1.448~~~&~~~1.529~~~ \\
  0.3& ~~~1.142~~~&~~~1.306~~~&~~~1.488~~~&~~~1.688~~~&~~~1.794~~~ \\
0.4& ~~~1.349~~~&~~~1.667~~~&~~~1.985~~~&~~~2.313~~~&~~~2.484~~~ \\
 0.5& ~~~2.520~~~&~~~3.620~~~&~~~4.621~~~&~~~5.608~~~&~~~6.108~~~ \\
 \hline
\end{tabular}
\mbox{}\\
\vspace{0.15in} Table I: Global mean number of images that are
minima (cf. Corollary~\ref{cor5}).
\end{center}

%\vspace{0.3in}

We see from the table that the global mean number of minima is
relatively small. For $\kappa_{tot} =  0.450$ and admissible
values of $\gamma$, $\kappa_c$ and $\kappa_*$, we observe that the
global mean number of minima is between $1$ and $3$.

Next we present values of $\widehat{E}[N_{g,min}(D)]$, the  global
mean number of  minima lying inside the disk $D=B({\bf 0},
1111.11)$. Here, we fix $g=1000$, $m=1$ and  $\kappa_{tot} =
0.450$ and we vary the values of $\kappa_*$ and $\gamma$.

%\vspace{0.3in}

\begin{center}
\begin{tabular}{|c|c|c|c|c|c|}\hline
\multicolumn{6}{|c|}{} \\
 \multicolumn{6}{|c|}{$\widehat{E}[N_{g,min}(D)]$:\  Global Mean Number of Minima in $D$ } \\
 \multicolumn{6}{|c|}{} \\
 \cline{1-6}
$ \gamma\diagdown\kappa_*$&~~~0.1~~~&~~~0.2~~~&~~~0.3~~~&~~~0.4~~~&~~~0.45~~~ \\
\hline
   0.1& ~~~1.036~~~&~~~1.113~~~&~~~1.217~~~&~~~1.343~~~&~~~1.412~~~ \\
   0.2& ~~~1.067~~~&~~~1.170~~~&~~~1.297~~~&~~~1.446~~~&~~~1.526~~~ \\
  0.3& ~~~1.142~~~&~~~1.305~~~&~~~1.486~~~&~~~1.685~~~&~~~1.791~~~ \\
0.4& ~~~1.349~~~&~~~1.666~~~&~~~1.982~~~&~~~2.309~~~&~~~2.479~~~ \\
 0.5& ~~~2.519~~~&~~~3.617~~~&~~~4.615~~~&~~~5.599~~~&~~~6.096~~~ \\
 \hline
\end{tabular}
\mbox{}\\
\vspace{0.15in} Table II: Asymptotics of the  global mean number
of   minima over the disk $B({\bf 0}, 1111.11)$
(cf.~Theorem~\ref{theo}).
\end{center}

\vspace{0.3in}

 The values in Table II differ from the corresponding  values
in Table~I by a relatively  small amount. This is expected as the
difference between  $\widehat{E}[N_{g,min}(D)]$ and ${\widehat
{E}} [N_{min}]$ has order $O(g^{-1})$ (by Theorem~\ref{theo} and
Corollary~\ref{cor5}).

\section{Conclusion}
We carried on our development of a mathematical theory for
stochastic microlensing, devoting our efforts to the study of the
expected number of different types of lensed images. This involved
an analysis of the random microlensing shear. In particular,  we
obtained the first three leading terms in their  asymptotic joint
p.d.f.s in the limit of an infinite number of stars  and showed
that each of their marginals converges to a shifted Cauchy
distribution, which has a heavy tail.  The third leading order
term of this joint p.d.f. depends on the magnitude of the shear
due to the  stars, the optical depth, and the mean number of stars
within a disk centered at the origin. We also obtained the
asymptotic p.d.f. of the shear's magnitude in the
 limit of an infinite number  of stars. These results on the random shear are given
for a general point in the lens plane.
Using the Kac-Rice formula and Morse theory, we determined the
expected total number of images and expected number of saddle
images for a general random lensing map. 
By formulating a generalization of the Kac-Rice type formula
to a family of compact subsets with positive measure of the light source plane,
we introduced the {\it global} expected number of positive parity images for
a general random lensing situation.  Our general results are applicable to a wide range of
lensing models. Specifically, we computed the global expected
number of minimum images for the case of microlensing. This global
expectation  was also illustrated with explicit numerical values.

\section{Acknowledgments}
We thank the referees for invaluable feedback that strengthened
the paper. AMT would like to thank A. Aazami, R. Adler, J.
Mattingly, and A. Watkins for helpful discussions. AOP
acknowledges the support of NSF Grants DMS-0707003 and
AST-0434277-02. Part of this work was done at the
Petters Research Institute in Belize.

\appendix

\section{Proof that \ $\mathfrak{I}_n \ = \ 0 .$}

We consider the integral
\[\mathfrak{I}_n\equiv \int_{R-|x|}^{R+|x|} \frac{\sin\left[n f(r)\right]}{r^{n-1}} \ dr\]
for any  integer $n\ \geq \ 2$ and   $0 \ \leq \ |x| \ < R,$ where
\[ f(r)= \arccos \left(\frac{|x|^2 \ + \ r^2 \ - \ R^2}{2 |x| r}\right).\]
Define the variable $\vartheta$ by
\[ R\cos \vartheta = r \cos f(r) - |x|
%= \frac{r^2 - R^2 - |x|^2}{2|x|}; \Rightarrow r^2= R^2 + |x|^2 + 2 R |x| \cos \vartheta
,\]
with $\vartheta$ restricted to the interval $[ 0 , \pi]$. This
implies that $\ r \sin f(r)\ = \ R  \sin \vartheta$. Therefore,
\beqan
r\exp \left[i f(r)\right]\ = \ |x| \ + \ R
e^{i\vartheta}\qquad {\rm and }\qquad r\exp \left[-i f(r)\right]\
= \ |x| \ + \ R e^{-i\vartheta}.
\eeqan
We now use these equalities and the change of variable from $r$ to
$\vartheta$  to obtain the following:
\beqn
 \mathfrak{I}_{n+1}&=& % \int_{R-|x|}^{R+|x|} \frac{\sin\left[(n + 1) f(r)\right]}{r^{n}} \ dr \ = \
\frac{1}{2i}\int_{R-|x|}^{R+|x|}\left[\frac{e^{i(n+1)f(r)}\ - \
e^{-i(n+1)f(r)}}{r^{n}}
   \right] dr\\
%%\frac{1}{2i}\int_{R-|x|}^{R+|x|}\left[\frac{e^{i(n+1)f(r)}-e^{-i(n+1)f(r)}}{r^{n+1}}
 %%  \ - \  \frac{e^{-i(n+1)f(r)}}{r^{n+1}}\right]\ r\ dr\\
&=& \frac{1}{2i}\int_{R-|x|}^{R+|x|}\left[\frac{1}{\left(r \,
e^{-if(r)}\right)^{n+1}}
   \ - \  \frac{1}{\left(r \, e^{if(r)}\right)^{n+1}}\right]\ r\ dr\\
&=& \frac{R |x|}{2i} \int _0^{\pi} \left[\frac{1}{\left(|x| \ + \
R e^{-i\vartheta}\right)^{n+1}}
   \ - \  \frac{1}{\left(|x| \ + \ R e^{i\vartheta}\right)^{n+1}}\right]\ \sin\vartheta \ d\vartheta\\
%&=& \frac{R |x|}{2i}  \left[ \int _0^{\pi}\frac{\sin\vartheta}{\left(|x| \ + \ R e^{-i\vartheta}\right)^{n+1}}d\vartheta
%   \ - \ \int _0^{\pi}\frac{\sin\vartheta}{\left(|x| \ + \ R e^{i\vartheta}\right)^{n+1}}d\vartheta\right] \\
%&=&  \frac{R |x|}{2i}  \left[ \int _0^{\pi}\frac{\sin\vartheta}{\left(|x| \ + \ R e^{-i\vartheta}\right)^{n+1}}d\vartheta
 %  \ + \ \int _{-\pi}^0 \frac{\sin\vartheta}{\left(|x| \ + \ R e^{-i\vartheta}\right)^{n+1}}d\vartheta\right] \\
&=& \frac{R |x|}{2i}   \int _{-\pi}^{\pi}\frac{\sin\vartheta}{\left(|x| \ + \ R e^{-i\vartheta}\right)^{n+1}}d\vartheta\\
&=& \frac{R |x|}{(2i)^2 R^{n+1}} \int _{-\pi}^{\pi}
\frac{e^{i\vartheta} \ - \ e^{-i\vartheta}}
{e^{-i(n+1)\vartheta}\left(1 \ + \ \frac{|x|}{R} e^{i\vartheta}\right)^{n + 1}} d\vartheta\\
&=& -\frac{|x|}{4R^n}\int_{-\pi}^{\pi} \left(e^{i(n + 2)\vartheta}
\ - \ e^{in\vartheta}\right)
\left(1 \ + \ \frac{|x|}{R} e^{i\vartheta}\right)^{-(n + 1)} d\vartheta\\
&=& \frac{|x|}{4R^n} \sum_{k=0}^{\infty} \frac{(-)^k
\Gamma(k+n+1)}{k! \Gamma(n+1)}\frac{|x|^k}{R^k}
   \int_{-\pi}^{\pi} \left(e^{i(n + k)\vartheta} \ - \ e^{i(n + 2 + k)\vartheta}\right)d\vartheta\\
&=& \ 0. \eeqn
Above, we used the inequality $0\ \leq \ |x| \ < \ R$ to obtain
the Taylor series expansion. Also, the convergence of the series
allowed us to reverse the order of  the summation  and the
integral. Finally, the last line follows by direct integration,
where we used the assumption $n \, + \, 1 \ \geq \, 2$, that is,
$n \,  \geq \, 1$.

%The integral $ \int_{-\pi}^{\pi} \ e^{i m \vartheta} \ = \ 2\pi \,
%\delta_{m,0}$ for any integer $m$. Thus, $ \int_{-\pi}^{\pi}
%\left(e^{i(n + k)\vartheta} \ - \ e^{i(n + 2 +
%k)\vartheta}\right)d\vartheta \ = \ 0 $ for any nonnegative
%integer $n$ and positive integer $n$. Hence,
%\[\mathfrak{I}_{n+1} \ = \ 0  \]
%for any positive integer $n$.

\end{document}